\documentclass[sigconf]{acmart}

\usepackage{mdframed}

\mdfsetup{skipabove=\topskip,skipbelow=\topskip}
\mdfdefinestyle{poistyle}{%
    frametitlerule=true,
}
\mdtheorem[style=poistyle]{poi}{Improvement}

\AtBeginDocument{%
  \providecommand\BibTeX{{%
    \normalfont B\kern-0.5em{\scshape i\kern-0.25em b}\kern-0.8em\TeX}}}

\setcopyright{none}
\renewcommand\footnotetextcopyrightpermission[1]{}

\begin{document}

\title{Multilingual Crowd-Based Requirements Engineering Using Large Language Models}

\pagestyle{empty}

\author{Arthur Pilone}
\email{arthurpilone@usp.br}
\orcid{0009-0004-3899-4087}
\affiliation{%
    \institution{University of São Paulo}
  \country{Brazil}
}

\author{Paulo Meirelles}
\email{paulormm@ime.usp.br}
\orcid{0000-0002-8923-2814}
\affiliation{%
    \institution{University of São Paulo}
  \country{Brazil}
}

\author{Fabio Kon}
\email{kon@ime.usp.br}
\orcid{0000-0003-3888-7340}
\affiliation{%
    \institution{University of São Paulo}
  \country{Brazil}
}

\author{Walid Maalej}
\email{walid.maalej@uni-hamburg.de}
\orcid{0000-0002-6899-4393}
\affiliation{%
  \institution{Universität Hamburg}
  \country{Germany}
}

\renewcommand{\shortauthors}{Pilone et al.}

\begin{abstract}
A central challenge for ensuring the success of software projects  is to assure the convergence of developers’ and users’ views. 
While the availability of  large amounts of user data from social media, app store reviews, and support channels bears many benefits, it still remains unclear how software development teams can effectively use this data. 
We present an LLM-powered approach called DeeperMatcher that helps agile teams use crowd-based requirements engineering (CrowdRE) in their issue and task management. We are currently implementing a command-line tool that enables developers to match issues with relevant user reviews. We validated our approach on an existing English dataset from a well-known open-source project. 
Additionally, to check how well DeeperMatcher works for other languages, we conducted a single-case mechanism experiment alongside developers of a local project that has issues and user feedback in Brazilian Portuguese. 
Our preliminary analysis indicates that the accuracy of our approach is highly dependent on the text embedding method used. 
We discuss further refinements needed for reliable crowd-based requirements engineering with multilingual support.
\end{abstract}


\maketitle

\section{Introduction}

Two key principles of agile software development are (a) to structure the work in short cycles, each with a set of tasks and issues to be resolved within that cycle, and (b) to repeatedly collect and prioritize feedback from the users and stakeholders \cite{Martens:Software:19}.
To identify and analyze requirements, agile methods mainly rely on the concept of \textit{user stories}, which forces development teams to think about the users performing a given task. 
Each story describes how a hypothetical user would use a relevant feature of the software system and why~\citep{beck2004xp, agile-practices}.

Although valuable for development teams and still considered effective for capturing user requirements, writing user stories may not be feasible for every situation and every type of requirement. 
Moreover, having a skilled customer or a product owner write good user stories is often far from reality for many software projects. 
It is thus common to see teams just keeping track of the tasks to be implemented and issues to be resolved \cite{Montgomery:MSR:2022, Montgomery:NLP4RE:2024}. 
The convenience of focusing on recording development and maintenance tasks comes with the risk that the development tasks  and the actual needs of the users may drift apart.

This scenario may be even more challenging for smartphone apps with large user bases \cite{Pagano2013}. 
A specific requirement related to a specific group may never get to the development team due to the absence of direct communication between developers and users.
It is not hard to see how the problem is exacerbated when it comes to bug  reporting. 
The vast range of operating systems, device models, execution environments, and user contexts make the task of accurately understanding how a given software product should behave in every possible condition nearly unattainable \cite{Martens:RE:2019, Maalej:NLP4RE:2024}.

With thousands of user reviews and comments on app stores and social media, prioritizing which comments should be analyzed by the development team to capture the requirements is challenging and demands an unreasonable amount of manual effort. 
The approach of tapping into the relevant  information instilled in these reviews and posts on social media has been named \textbf{Crowd-Based Requirements Engineering} (CrowdRE)~\citep{CrowdRE17, CrowdCenteredRE14}. 
This approach has been studied before, especially alongside the use of machine learning and natural language processing (NLP) \citep{Santos2019AnOO, Stanik2019, RohanClassify}, with a variety of datasets, use-cases, and levels of automation accuracy \cite{Maalej:NLP4RE:2024}. 

The rise of increasingly powerful large language models (LLMs) opens the door to many new applications revolving around summarizing, understanding, and generating natural language. 
The rapid emergence of this technology fosters research on its possible use cases in requirements engineering \citep{borg2024} and software engineering in general \citep{fan2023Preprint}.
To leverage the ever-evolving potential of LLMs for crowd-based requirements engineering, we are developing an automated approach for assisting development teams identify user reviews corresponding to development tasks, or a lack thereof.

The remainder of the paper is structured as follows: 
In Section~\ref{sec:proposal}, we propose a high-level architecture for an approach that leverages LLMs to promote multilingual crowd-based requirements engineering. Section~\ref{sec:implementation} then reports on our ongoing open-source implementation for the proposed system.
In Section~\ref{sec:results}, we present a preliminary evaluation for our implementation based on an available English dataset and a new dataset collected from a Brazilian Portuguese project. We then discuss how our results point toward new directions to grow and evolve the proposed tool in Section~\ref{sec:discussion}. 
Finally,  we summarize our proposal and future directions in Section~\ref{sec:conclusion}.

\section{Approach}
\label{sec:proposal}

A common step for most Natural Language Processing (NLP) tasks is to split the input text into \textit{tokens} and assign a numeric representation to each one of them.
 This numeric representation can be condensed to an 
\textit{embedding}, a list of floating point numbers that stores semantic information relative to its corresponding token. 
Typical LLMs are based on the Transformer architecture~\citep{vaswani2017attention}, which promotes the exchange of information between embeddings of tokens from a single string. 
Interestingly, this interaction is heavily connected to the concept of \textit{attention}~\citep{kim2017} and is the basic idea behind most recent advances in NLP. 
As the models process text, the embedding of a token is influenced by the embeddings of its neighboring tokens, similar to how the meaning of a noun is influenced by the adjectives surrounding it. 
At the encoder output, each embedding carries information that reflects the token context. 
These final embeddings can be called  \textit{contextualized embeddings}. 
As these embeddings strongly correlate with the token meanings~\cite{jiao2021}, we can try to measure the semantic similarity of two texts by the proximity between their embeddings in their high-dimensional  vector space.

In this line, we propose a system that uses the \textbf{contextualized embeddings} created by LLMs to measure the similarity of user reviews and development issues and suggest matches of pairs that might point to the same topic. 
Development teams that receive large amounts of user feedback should be able to use our approach to identify what issues in their issue tracker a user review may correspond to. 
When the tool matches a user feedback item to a bug report from the issue tracker, the developers should be confident that the user is reporting something they are already aware of. 
On the other hand, when the tool does not find a suitable match for a user review, the development team should consider giving it a further inspection and possibly creating a new issue or task.

\begin{figure}[ht]
  \centering
  \includegraphics[width=0.75\linewidth]{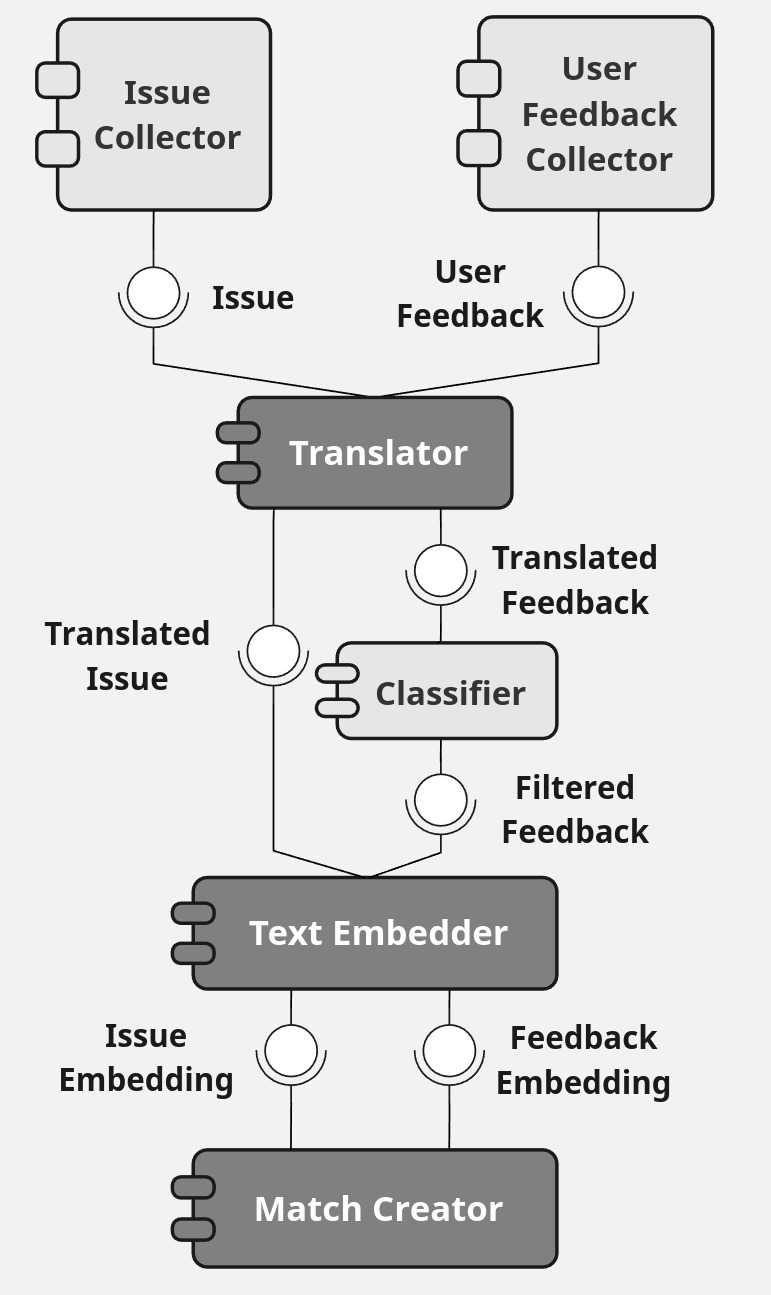}
  \caption{Core components of the proposed architecture. In a darker shade of gray, we highlight the components that receive data from both user reviews and issues. Every component depicted can be switched or adapted for the needs of specific teams or projects.}
  \label{fig:arch}
  \Description{Core components of the proposed architecture.}
\end{figure}

Figure~\ref{fig:arch} depicts the main components of our approach. 
A guiding principle in our design is to give developers the flexibility to switch and modify components that are likely to change due to rapidly evolving NLP techniques or specific team preferences. 

One of the key functions of the approach is to extract and collect issues and user reviews. 
The system architecture includes components that interact with external APIs and crawlers to gather this data. 
The components responsible for collecting user reviews and developer issues follow a straightforward interface, ensuring that future maintainers can easily implement classes to extract data from new sources. 

Besides, the abstractions created for issues and user feedback must be generic enough to avoid coupling to specific platforms and enable the desired flexibility. 
As the models used for review classification and text embedding may only work with text in a single language (English, for instance), we introduce a text-translator component used after the system collects the data. 
This is vital to support projects with reviews in multiple languages. 
However, the need for this component depends entirely on the choices made by the system maintainers.

Before embedding user reviews, a team of developers may choose to filter those that might be irrelevant~\citep{Stanik2019} or do not correspond to any of the issue types maintained in the repository. 
For instance, the team might not be interested in reviews requesting new features or simply praising the app. Accordingly, our architecture can also include a component for classifying user feedback \citep{Maalej:NLP4RE:2024}. 
Similar to the text translator, this component can be adapted depending on the review types of interest to the development team.

The text embedder is the part of the system responsible for receiving a text (i.e, a review or an issue) and returning a single embedding used to compare it to other texts. 
We provide the text to an LLM and use the contextualized embeddings it computes to generate a single embedding for the given string. 
The system creates an embedding for each review and developer issue.
As this component relies heavily on the specific LLM used, our architecture also allows for changing it according to the project's needs.

After computing the contextualized embeddings, the system measures the distance between them to estimate the similarity based on their proximity in the high-dimensional space \citep{mikolov2013} and suggests possible matches. 
Since the metric used for computing this distance is also subject to change, the system should support different similarity thresholds as well as adjustments in the number of issues suggested for each user review and vice versa.

Another guiding principle of our proposed approach  is to support the requirements engineering process while maintaining a human in the loop \citep{Andersen:Software:2024}. 
While leveraging LLMs, we aim to mitigate the risk of fully relying on the inherently imperfect nature of artificial intelligence predictions in the software development process. 
Future extensions to our architecture should be cautious to avoid introducing imprecise results that could jeopardize the development of software products and their users.

\section{Implementation}
\label{sec:implementation}

We are developing an open-source tool following the design decisions proposed in the previous section. Our system, nicknamed \texttt{DeeperMatcher}, is based on the text embedding approach of \textit{DeepMatcher} \cite{haering2021}. 
It is a command-line interface written in Python and is available at the GitLab repository \url{https://gitlab.com/ArthurPilone/deepermatcher}.

Currently, the system includes an issue collector for public GitLab repositories and stubs for other crawlers and collectors inherited and adapted from the DeepMatcher replication package. 
We initially chose GitLab due to its widespread adoption for hosting open-source projects and because it hosts the sample project used for our preliminary evaluation. 
We anticipate implementing extensions for these components in the near future. 
For our implementation of issue collectors, we follow the findings from ~\citet{haering2021} and use only the issue titles for textual embedding, as they have been shown to adequately summarize the issue content.

To support languages other than English, we implemented a simple text-translator component using the \texttt{googletrans} API, which is built on Google Translate. 
This component can translate text between any two languages supported by the API. 
\texttt{DeeperMatcher} currently translates issues to English during collection and translates user reviews after they are entered in the command-line interface. 
This class can be replaced with other text-translator components if users prefer to use alternative translation service.

Similar to DeepMatcher, we used the classifier from ~\citet{Stanik2019} to classify user reviews into three classes: ``Irrelevant'', ``Feature Request'', or ``Bug Report''. 
Using their replication package, we trained a simple classifier built on the DistilBERT~\citep{sanh2019distilbert} model from the \texttt{Transformers} Python library. 
If desired, a user of \texttt{Deeper} \texttt{Matcher} can filter reviews to only match bug reports.

For the text embedding process, \texttt{DeeperMatcher} currently features two approaches. 
The first mirrors DeepMatcher. 
We feed the input text to a DistilBERT model, which tokenizes it and generates a numeric embedding for each token. 
The model adjusts each token’s embedding based on the embeddings of neighboring tokens, so that every embedding also carries information about the context in which the token appears. 
Next, we use the SpaCy~\citep{spacy} part-of-speech (POS) tagger to identify tokens related to nouns and use the pytokenizations module to align the tokens from the SpaCy and DistilBERT models. 
Finally, we collect the contextualized embeddings generated by DistilBERT for the tokens that the SpaCy model identified as nouns. 
To create a single embedding for the entire text, we compute the average of the embeddings of all tokens in the text. 

The second embedding approach uses the \texttt{SentenceTransformers} \citep{reimers2019sbert} Python library. The model, ``\texttt{all-MiniLM-L6-v2},'' and the library are both designed to compute a single meaningful embedding for a given text string, thus providing an equivalent approach to the text embedding process used in DeepMatcher.

Following the approach from DeepMatcher, we use cosine similarity to quantify the similarity between issue and review embeddings. 
The user may specify a similarity threshold, and the number of matches \texttt{DeeperMatcher} should suggest for each user review.

The next step in developing \texttt{DeeperMatcher} is to incorporate new approaches for text embeddings using newer LLMs. 
Notably, we will adapt the current system to facilitate the transition between text embedding methods and introduce new options beyond those currently available. 
Additionally, the tool will receive usability improvements soon, including new options for outputting results, new data collectors, and the suggestion of issue titles when \texttt{DeeperMatcher} does not find a matching entry for a given review.

\section{Preliminary Evaluation}
\label{sec:results}

Trust should define the relationship between the proposed tool and the teams using it. If \texttt{DeeperMatcher} does not suggest an existing issue for a user review, the developer must be able to trust the system and assume that there is no relevant issue in their issue tracker. However, if \texttt{DeeperMatcher} frequently fails to suggest an existing issue for a matching review, the developer may lose trust in the tool, diminishing its value. Therefore, it is crucial that the system consistently identifies pairs of inputs that describe a common feature or problem. We refer to this property as \textit{reliability}.

\subsection{Evaluation Goal}

Our evaluation investigates how well the approach works in its current implementation and identifies potential areas for improvement.
 We focus on answering the following question:
 
\begin{itemize}
    \item {How reliable are the matches suggested by \texttt{DeeperMatcher} for English and other languages?}
\end{itemize}

\texttt{DeeperMatcher} builds upon the DeepMatcher proof of concept. Therefore, it is expected that \texttt{DeeperMatcher} functions properly with the issues and reviews used to validate the original proof. A fundamental principle in the development of \texttt{DeeperMatcher} was to ensure its applicability to data compatible with DeepMatcher.

\begin{figure}[ht]
  \centering
  \includegraphics[width=\linewidth]{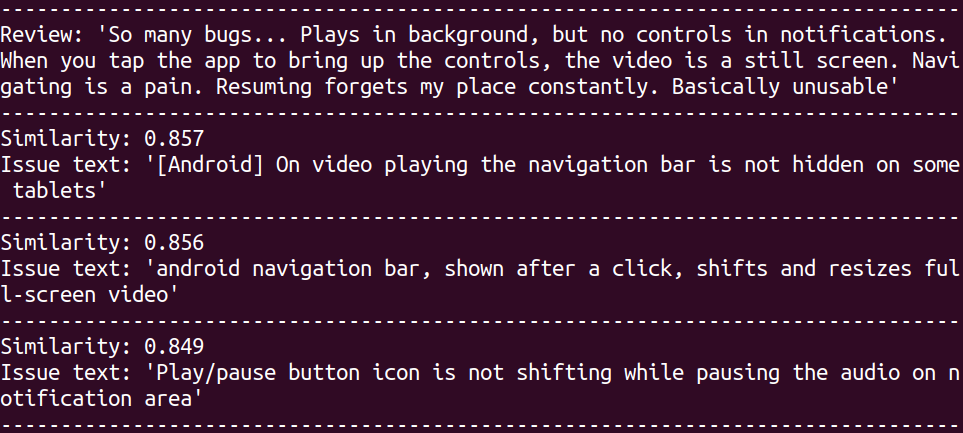}
  \caption{Screenshot of matches identified by \texttt{DeeperMatcher} when prompted with a review from Table III of the DeepMatcher proof of concept \citep{haering2021}.}
  \label{fig:og_match}
  \Description{Screenshot of matches found by \texttt{DeeperMatcher} when prompted with a review from Table III of the DeepMatcher proof of concept \citep{haering2021}.}
\end{figure}

Using the command-line interface of \texttt{DeeperMatcher}, it is easy to verify that the new architecture produces matches equivalent to those from the DeepMatcher proof of concept. Figure~\ref{fig:og_match} illustrates an example where \texttt{DeeperMatcher} results coincide with those from DeepMatcher. We anticipate that the text embedding generation in \texttt{DeeperMatcher} will improve and be easily adjustable. Therefore, our evaluation does not focus on the direct characteristics of the numeric values created for the text embeddings. Instead, we concentrate on assessing the reliability of \texttt{DeeperMatcher} as a tool.

\subsection{Data Acquisition}

To compare the performance of \texttt{DeeperMatcher} with its predecessor, we use the same data that was used to evaluate DeepMatcher. This allows us to observe changes in the matches suggested by both systems. We utilized data from~\citet{haering2021}, which includes English-written issues and reviews from two large-scale open-source projects: the VLC media player and the Signal messaging app. We selected data from these projects for our comparison.

Additionally, to evaluate the reliability of \texttt{DeeperMatcher} with data in languages other than English, we collected 574 issues and 69 user reviews in Brazilian Portuguese from the medium-scale project BikeSP \citep{PiloneBikeSP}. The BikeSP app developers manually associated each review with the corresponding issue when the user comment referred to something identifiable solely by its text. Out of the 69 reviews, only 23 had a corresponding development issue. Our replication package also includes the reviews and issues from this project.

\subsection{Evaluation Methodology}

First, we test whether the text embedder implementation derived from DeepMatcher produces results equivalent to those of the original prototype. We provide 100 randomly selected user reviews from the Signal Messenger app data to \texttt{DeeperMatcher} and verify if the three issues suggested by the tool match those listed by DeepMatcher for the same input.

Next, we focus on the main part of our evaluation. We conduct a single-case mechanism experiment~\citep{Wieringa2014SingleCase} to study how \texttt{Deeper} \texttt{Matcher} performs on data from the BikeSP and VLC projects and how variations in input affect the results. Using the command-line interface (CLI) of \texttt{DeeperMatcher}, we set the source of issues to the repository for the desired project and instruct the tool to translate user feedback from Brazilian Portuguese to English. We also disable the minimum similarity threshold and configure the system to provide five candidates for matching issues for each review.

We enter the reviews one by one and analyze the results individually. For reviews where the Brazilian developers identified a matching issue, we count how frequently \texttt{DeeperMatcher} lists the issue the developers had in mind. This procedure is repeated twice for each user review from the Brazilian project: first using the text embedding method from DeepMatcher and then using our newer text embedding method based on \texttt{SentenceTransformers}.

Finally, we conduct a qualitative analysis by selecting issues from the VLC and BikeSP projects and inserting corresponding or newly created reviews. We observe how the matches suggested by \texttt{DeeperMatcher} vary with different text embedding methods.

\subsection{Quantitative Results}

Our test suite confirmed that our implementation is consistent with the original proof of concept from which it was derived. For all 100 reviews sampled from the Signal app data, \texttt{DeeperMatcher} consistently suggested the same three issues as DeepMatcher.

However, when tested with the text embedding method derived from DeepMatcher, \texttt{DeeperMatcher} struggled with most reviews. Out of 29 reviews where the Brazilian developers had identified a matching issue, only three (13.0\%) were included by \texttt{DeeperMatcher} in its list of the five most similar issues. In these cases, the mean similarity value for the correct issue was 81\%, with two issues listed as the 5th most similar and one as the 2nd most relevant.

In contrast, results improved significantly with the newer embedding method. Using the LLM from the \texttt{SentenceTransformers} library, \texttt{DeeperMatcher} correctly suggested the issue for 13 out of 23 (56.5\%) reviews. Notably, most of these matches had a similarity value of less than 80\%, indicating effective but modest alignment.

These results demonstrate that the accuracy of matches is highly dependent on the text embedding method used. They also suggest that further improvements to \texttt{DeeperMatcher} may be needed to achieve higher prediction reliability.

\subsection{Qualitative Results}

\begin{figure}[ht]
  \centering
  \includegraphics[width=\linewidth]{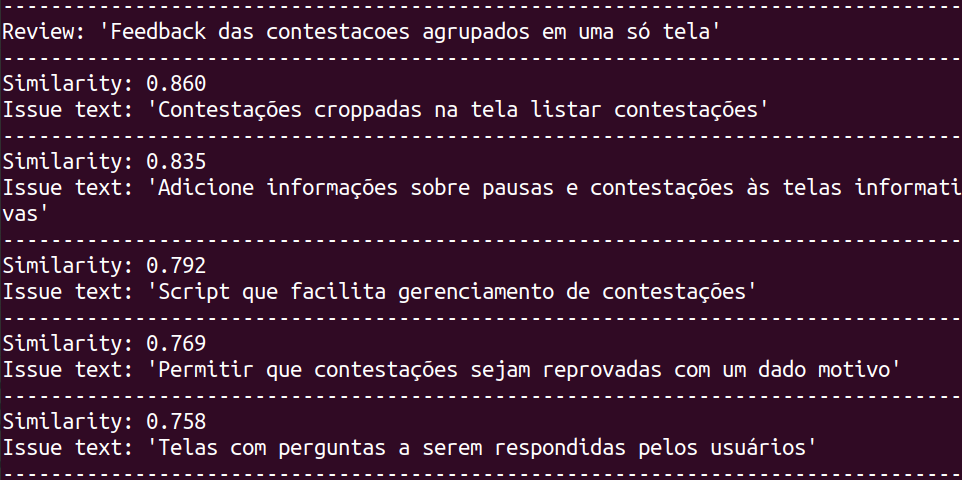}
  \caption{Suggested issues for a review requesting a new app screen: Although the issue related to the creation of the new screen is not listed, the first suggested match is a fix for a problem with the existing screen.}
  \label{fig:fix_mentioned_match}
  \Description{Suggested issues for the review requesting a new screen for the app.}
\end{figure}

We noted and investigated some interesting patterns as we experimented with the embedding method inspired by DeepMatcher. Although many suggestions did not correspond exactly to the issue representing the implementation of a new feature, they often included issues describing fixes or extensions related to the feature. For instance, as illustrated in Figure~\ref{fig:fix_mentioned_match}, \texttt{DeeperMatcher} did not identify the issue describing the creation of a new screen for the app. However, it suggested another issue with high similarity (86\%) that described a problem with the screen.

Furthermore, we observed that the original text embedding method performed poorly when user reviews were \textit{considerably longer than the corresponding issue summary}. The additional text created noise that interfered with its contextualized 

To analyze the relationship between feedback length and accuracy of the match, we experimented with forging user reviews of different lengths describing requests and reports for problems present in the issue database. For the issue ``Impedir início de viagem se economia de bateria estiver ligada'' [``Stop users from starting a trip if the battery saver is on''], the review ``Não existe uma maneira de impedir que eu inicie uma viagem se eu estiver com economia de bateria?'' [``Isn't there a way to stop me from starting a trip if I have the battery saver on?''] leads to match with an 83.1\% similarity score, but ``Por que não me impedem de iniciar uma viagem se sabem que eu estou com economia de bateria?'' [``Why don't they stop me from starting a trip if they know I have the battery saver on?''] does not lead to a satisfying match.

Surprisingly, this problem is not exclusive to translated issues. In the VLC dataset, we observed that the issue ``Audio cuts off on Android'' is the third suggestion \texttt{DeeperMatcher} provides when prompted with the review ``The audio keeps cutting off,'' with an 80\% similarity rating. However, altering the review text to ``I don't understand why the audio keeps getting cut off'' causes \texttt{DeeperMatcher} not to list the corresponding issue among the first ten matches.

Repeating these two tests with the newer embedding method, we observe signs of improvement. Although the examples in Portuguese yield the same results for the \texttt{SentenceTransformers} embedder, we were unable to create a user review containing both ``audio'' and ``cuts off'' without \texttt{DeeperMatcher} suggesting ``Audio cuts off on Android'' as a matching issue.

Our evaluation shows that \textbf{\texttt{DeeperMatcher} is still unreliable in its current state}. Further improvements are necessary before the tool can be applied in practice. It appears that the older text embedding approach performed poorly with the Brazilian Portuguese data, failing to identify the relevant issues manually identified by the developers. Additionally, we have identified a new potential point of failure in how \texttt{DeeperMatcher} handles longer user reviews.

\section{Discussion}
\label{sec:discussion}

From our preliminary evaluation, we conclude that several possible improvements to \texttt{DeeperMatcher} should be explored before it can deliver reliable matches for any project in any language. Our results are somewhat less favorable than those reported by~\citet{haering2021}. However, this does not contradict the findings from previous work, nor does it necessarily indicate degraded performance with the new system architecture. Nevertheless, we reflect on our evaluation results to identify potential \textbf{improvement points} for \texttt{DeeperMatcher} and its architecture.

First, our methodology differs from that used to evaluate the DeepMatcher proof of concept. While its authors focused primarily on identifying matches with bug reports, we expanded the scope to include reviews and issues related to new features. Additionally, our evaluation method diverges critically from DeepMatcher's approach. By having access to the development team, we collected precise information on which issues were created in response to each user review. This valuable data allowed us to conduct a more rigorous evaluation. Instead of merely counting matches considered relevant by coders, we quantified exact matches identified by individuals actively involved in the project.

Additionally, we should reflect on the differences between the new data we are using and the datasets used for the DeepMatcher evaluation. Setting aside the language aspect, the project analyzed in our study is critically distinct from those in the DeepMatcher evaluation. Although the number of issues from the Bike SP project is comparable to that of the VLC project, the issues from the Brazilian team include not only feature requests and bug reports but also management tasks. Moreover, these issues are frequently interrelated, as features implemented by the team are often expanded or require new fixes. This high interconnectivity increases the similarity among different issues in the repository, making it challenging for the review embeddings to be distinctly closer to a single issue.

Therefore, we argue that \texttt{DeeperMatcher} can exhibit variable performance depending on the issue repository. Implementing additional pre-processing or pre-selection steps between issue collection and its use for suggestions may be crucial for achieving more accurate matches.

\begin{poi}
It might be necessary to \textbf{cluster issues based on the common features} they refer to or to \textbf{exclude issues that might be excessively technical or irrelevant} to what users can review. This step could be achieved through manual filtering by the development team (e.g., by adding an additional issue field) or by integrating a new dedicated component into the architecture of \texttt{DeeperMatcher}.
\end{poi}

When checking the influence of review length on the resulting text embedding, we observe a limitation inherent to contextualized embeddings and LLMs in general. As newer and more powerful LLMs are developed, their high-dimensional textual embeddings tend to improve the matching metrics significantly. The improvement in results after we changed the text embedding highlights how more recent embedders can mitigate this problem. Therefore, we conclude that constantly updating the models used for the text embedding process should enhance the matching performance.

\begin{poi} Switching DistilBERT for the original BERT or a larger model like Meta's Llama3 should be a straightforward upgrade due to our proposed architecture. Additionally, leveraging our system's adaptability to \textbf{include LLMs specifically built for clustering similar texts}, such as those from \texttt{Sentence} \texttt{Transformers}, can further improve the text embedding process.
\end{poi}

Another way to mitigate the problem with imprecise embeddings in longer texts is to add a pre-processing step before the embedding process. The goal is to select only the most relevant parts of the review or issue text, thereby preemptively de-noising the text.

\begin{poi}
    Using another review processing component \citep{Maalej:NLP4RE:2024}, one might \textbf{delineate the most essential parts of each review}. Alternatively, a review processing component could \textbf{cluster reviews or issues} before \texttt{DeeperMatcher} searches for matches.
\end{poi}

Unsurprisingly, translating the input text before creating the embedding negatively impacts the matching performance. The translation process can introduce slight changes in meaning, as different nuances may be ignored or lost. Feeding longer text entries to the translator increases the likelihood of these linguistic losses affecting the embedding. The added noise from translations also contributes to why \texttt{DeeperMatcher} struggled with the Brazilian Portuguese dataset. In addition to being in a different language, our reviews were, on average, longer than those in the DeepMatcher evaluation dataset, averaging 38 words per review compared to 33 in the original dataset \citep{haering2021}.

\begin{poi}
    \textbf{By incorporating text embedding methods that work directly with the languages used by developers and their users}, such as Brazilian Portuguese in our sample project, the need for a text translator component could be eliminated. Reducing this complexity might enhance the performance of \texttt{DeeperMatcher}.
\end{poi}

When comparing the results with our approach, clear evidence supports one of our core design choices: many identified improvements can be implemented without major restructuring of \texttt{Deeper} \texttt{Matcher}, thanks to the easy interoperability of its fundamental components, particularly the text translator and the text embedder.

We can take this discussion further by reflecting on the importance of continuous research in areas as rapidly evolving as LLMs and their applications in software engineering. Our results demonstrate how higher expectations and more advanced models can render recently presented solutions nearly obsolete. Therefore, continuous adaptability should be a guiding principle in designing systems based on LLMs.

\section{Conclusion}
\label{sec:conclusion}

We presented the design, implementation, and preliminary evaluation of \texttt{DeeperMatcher}, an innovative system leveraging the growing capabilities of LLMs for crowd-based requirements engineering. This system utilizes text embeddings created by LLMs to achieve a previously unfeasible task: matching relevant feedback from large volumes of user data to corresponding issues in the issue tracker. Its extensible architecture supports various models and the specific matching preferences of development teams.

\texttt{DeeperMatcher} is being developed as an open-source command-line utility that enables developers to identify matches between issues and user reviews. We conducted a single-case mechanism experiment with a medium-scale real-world project to evaluate its matching reliability. Our preliminary results indicate that further modifications are needed to provide reliable assistance to development teams globally. We specifically highlighted the need for an embedding based on a more powerful LLM than DistilBERT and for a filtering or pre-processing step \citep{Maalej:NLP4RE:2024} of issues before embedding. Our work underscores the importance of ongoing research into the effective use of LLMs for software and requirements engineering scenarios. Researchers should continuously revisit and reevaluate recent work involving LLMs as newer models and techniques emerge.

\section{Artifacts Availability}

The replication package for this article is publicly available at \url{https://zenodo.org/doi/10.5281/zenodo.11424209}. We provide A) the \\ \texttt{DeeperMatcher} source code as used in our analysis; B) the data used to train the classification model; and C) the data (from the projects) used in our analysis. We compressed each directory from the project repository, and they are available as zipped files. The README file contains basic information on the repository structure.

\begin{acks}
The authors of this paper have received financial support from FAPESP under the research grant 2024/00957-8.
\end{acks}

\bibliographystyle{ACM-Reference-Format}
\bibliography{main}

\end{document}